\documentclass{article}

\usepackage{graphicx}
\usepackage{natbib}

\usepackage[final]{neurips_2023}

\usepackage[utf8]{inputenc} % allow utf-8 input
\usepackage[T1]{fontenc}    % use 8-bit T1 fonts
\usepackage{hyperref}       % hyperlinks
\usepackage{url}            % simple URL typesetting
\usepackage{booktabs}       % professional-quality tables
\usepackage{amsfonts}       % blackboard math symbols
\usepackage{nicefrac}       % compact symbols for 1/2, etc.
\usepackage{microtype}      % microtypography
\usepackage{xcolor}         % colors
\usepackage[nameinlink]{cleveref}

\crefname{figure}{Fig.}{Figs.}
\Crefname{figure}{Figure}{Figures}

\title{Representational constraints underlying similarity between task-optimized neural systems}

\hypersetup{
    colorlinks = True,
    allbordercolors = {white},
    urlcolor = {cyan},
    citecolor = {gray},
    linkcolor = {gray}
}

\usepackage{etoolbox}
\makeatletter

\pretocmd{\NAT@citex}{%
  \let\NAT@hyper@\NAT@hyper@citex
  \def\NAT@postnote{#2}%
  \setcounter{NAT@total@cites}{0}%
  \setcounter{NAT@count@cites}{0}%
  \forcsvlist{\stepcounter{NAT@total@cites}\@gobble}{#3}}{}{}
\newcounter{NAT@total@cites}
\newcounter{NAT@count@cites}
\def\NAT@postnote{}

\def\NAT@hyper@citex#1{%
  \stepcounter{NAT@count@cites}%
  \hyper@natlinkstart{\@citeb\@extra@b@citeb}#1%
  \ifnumequal{\value{NAT@count@cites}}{\value{NAT@total@cites}}
    {\ifNAT@swa\else\if*\NAT@postnote*\else%
     \NAT@cmt\NAT@postnote\global\def\NAT@postnote{}\fi\fi}{}%
  \ifNAT@swa\else\if\relax\NAT@date\relax
  \else\NAT@@close\global\let\NAT@nm\@empty\fi\fi% avoid compact citations
  \hyper@natlinkend}
\renewcommand\hyper@natlinkbreak[2]{#1}

\patchcmd{\NAT@citex}
  {\ifNAT@swa\else\if*#2*\else\NAT@cmt#2\fi
   \if\relax\NAT@date\relax\else\NAT@@close\fi\fi}{}{}{}
\patchcmd{\NAT@citex}
  {\if\relax\NAT@date\relax\NAT@def@citea\else\NAT@def@citea@close\fi}
  {\if\relax\NAT@date\relax\NAT@def@citea\else\NAT@def@citea@space\fi}{}{}

\makeatother

\author{Tahereh Toosi\\
Center for Theoretical Neuroscience\\
Zuckerman Mind Brain Behavior Institute\\
Columbia University\\
New York, NY \\
\texttt{tahereh.toosi@columbia.edu} \\
}

\begin{document}

\maketitle

\begin{abstract}
Neural systems, artificial and biological, show similar representations of inputs when optimized to perform similar tasks.
In visual systems optimized for tasks similar to object recognition, we propose that representation similarities arise from the constraints imposed by the development of abstractions in the representation across the processing stages. To study the effect of abstraction hierarchy of representations across different visual systems, we constructed a two-dimensional space in which each neural representation is positioned based on its distance from the pixel space and the class space. Trajectories of representations in all the task-optimized visual neural networks start close to the pixel space and gradually move towards higher abstract representations, such as object categories. We also observe that proximity in this abstraction space predicts the similarity of neural representations between visual systems. The gradual similar change of the representations suggests that the similarity across different task-optimized systems could arise from constraints on representational trajectories.

\end{abstract}

\section*{Introduction}
Recent progress in neuroscience and AI has revealed that biological and artificial learning systems tend to form similar representations at various stages of processing when exposed to similar stimuli. This similarity has been observed in different modalities such as vision \citep{Yamins2014-ii, Khaligh-Razavi2014-gu, Schrimpf2020integrative}, auditory \citep{Kell_2018}, and language \citep{Schrimpf2021-fp}. This similarity in representations, particularly the similarity between intermediate levels of processing, has led to optimism about the potential of such similarities, between artificial systems and the brain, to help in understanding the mechanisms of brain function. However, as models with diverse objective functions and architectures are examined for these similarities, doubts arise regarding the inferences of mechanistic similarity in neural processing: If representational similarity indicates that the neural mechanisms underlying those representations are similar, then why is a comparable level of similarity observed despite significant structural and optimization differences \citep{SchrimpfKubilius2018BrainScore, Conwell_2022}? \\

In this work, we investigate the similarity of representations between biological and artificial visual systems that are optimized for tasks similar to object classification, such as recognition, localization, and segregation. To study the neural representations between visual systems, as representations are developed across processing stages, we introduce an abstraction space. This is a conceptual framework that allows measurement of the similarity of each representation both from the input (pixel) and output (class) representations. This abstraction space has two coordinates: on one axis, we measure the similarity of the representation relative to the pixel space and on the other axis, we measure the similarity of the representation relative to the object class space (\cref{fig:fig1}A). We compare the trajectories of neural representations across processing stages for different visual systems in this abstraction space between the pixel input and the category output. We also examine how the proximity in this two-dimensional space is related to the similarity of representations between visual systems. The results of this study suggest that similar underlying constraints on the abstraction trajectory lead to the similarity of representations between visual systems.

\section*{Results}
% First, we define the abstraction space and then examine the trajectories of different visual systems in this space.

\subsection*{Abstraction space}
Similarity between neural representations is measured when the networks receive the same stimuli. Here, we focus on a widely used stimulus set to study the similarity between biological vision and deep neural networks for object recognition \citep{Majaj2015-ax,Yamins2014-ii}. This stimulus set consists of 5760 grayscale images constructed from 64 objects from 8 categories with varying positions, views, and sizes superimposed on random natural backgrounds \citep{Majaj2015-ax,Yamins2014-ii,Hong2016-fn} (see Figure \ref{fig:fig1}A for an example of a rotated face on top of a natural background). The pixel space is defined by the pixel values of the images, and the class space is defined by one-hot vectors representing each category. We used Centered Kernel Analysis (CKA, \cite{Kornblith2019-gi}) to compute the similarity between each representation and these two spaces—pixel and class. However, similar results are obtained using other representational similarity metrics, such as Representation Similarity Analysis (RSA).\\

Independent of neural representations, we first assess the abstraction hierarchy in this abstraction space for features extracted directly from the image set (e.g., object position, background, etc.). After extracting each feature, we construct a corresponding feature space within the image data set \citep{toosi_ccn}. For example, the space of the object positions (detection in \ref{fig:fig1}A) is constructed by retaining only the pixels within a circular area around the object in each image. Similarly, the space of the segmented objects (segmentation in \ref{fig:fig1}A) is defined by keeping the pixels inside the object boundary. We then measure the similarity of each feature space to both the original pixel space and the class space (Figure \ref{fig:fig1}A). As expected, with this approach, we capture the progression of the abstraction for the features directly extracted from the images. This approach captures an array of both the high-dimensional features (e.g., pixels, edges, low-frequency or high-frequency parts of the image) and the low-dimensional features (e.g., detection labels, segmentation labels, object identification, and classification). When situated in this abstraction space, these features form a hierarchy of abstraction. For instance, background representations are less similar to the class space than to the pixel space, and detection (object position) is less similar to class representations than segmentation (object boundary).  \\

\subsection*{Visual systems have similar trajectories in the abstraction hierarchy}

\begin{figure}[!ht]
    \centering
    \includegraphics[width=1\linewidth]{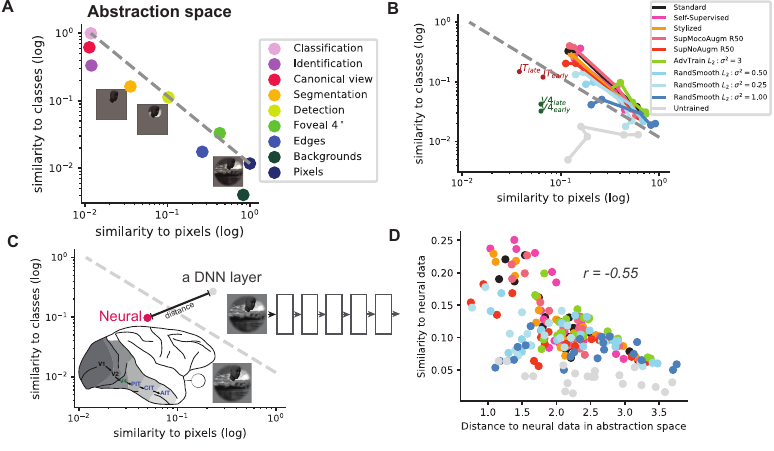}
    \caption{\textbf{A)} Similarity of each low-level and high-level feature spaces in images (e.g. edges and segmentation) to pixel space (x-axis) and to class space (y-axis), adapted from \citep{toosi_ccn}. \textbf{B)} Neural data recorded from the V4 and IT cortex of monkeys \citep{Majaj2015-ax, Yamins2014-ii} located in the abstraction space with green and red dots. Early and late denote the response latency after stimulus onset. Each trajectory shows the representations located in the abstraction space, for 5 layers of ResNet50 (layer1 to layer5 and avgpool), when trained under different objectives. \textbf{C)} The schematic of the Euclidean distance between each DNN layer and each neural data in the abstraction space.  \textbf{D)} The x-axis shows the distance between each biological neural data and each layer in B in the abstraction space. The y-axis shows the similarity of each biological neural data and each DNN layer in B measured with CKA. Similar colors as in B are used to show the similarity of each DNN network layer with the biological neural data.}
    \label{fig:fig1}
\end{figure}

To study the development of abstraction across different visual systems, we mapped both recorded neural activations from monkey visual cortices and DNN activations onto the defined abstraction space. The neural activity of the monkeys was recorded in the V4 and IT cortex. We included two timestamps for these recordings: early and late (for V4: 75ms and 105ms; for IT: 105ms and 145ms after stimulus onset, respectively). For DNNs, we used a family of ResNet50 architectures trained under various objective functions. These included standard classification \citep{resnet}, self-supervised, stylized \citep{geirhos2018}, adversarially robust \citep{robustness}, robust to Gaussian noise \citep{Certified_robust}, and some custom optimized with varying degrees of data augmentation. Across all these systems, the trajectory of representations across layers begins near the pixel space representation and gradually follows a similar abstraction hierarchy toward more abstract representation culminating in classification (Figure \ref{fig:fig1}B).\\

\subsection*{Proximity in the abstraction space predicts the representational similarity}
We evaluated how well proximity in the abstraction coordinate system predicts the pairwise similarity of the representations. To assess this, we compared the Euclidean distance in the abstraction space between each DNN layer representation and the biological neural recordings (Figure \ref{fig:fig1}C) against their pairwise representational similarity measured by CKA. The pairwise distance in the abstraction coordinate system is strongly correlated with the pairwise representational similarities (Pearson correlation, r=-0.55, p=1.9e-17, Figure \ref{fig:fig1}D). Therefore, the proximity of two representations in the abstraction coordinate system serves as a reliable proxy for their representational similarity. 

% Next, we show that the trajectory of representations for each optimized neural network in this coordinate system is constrained, thus enforcing the trajectories to be close to each other, giving rise to the similarity of representations.

\begin{figure}[ht]
    \centering
    \includegraphics[width=1\linewidth]{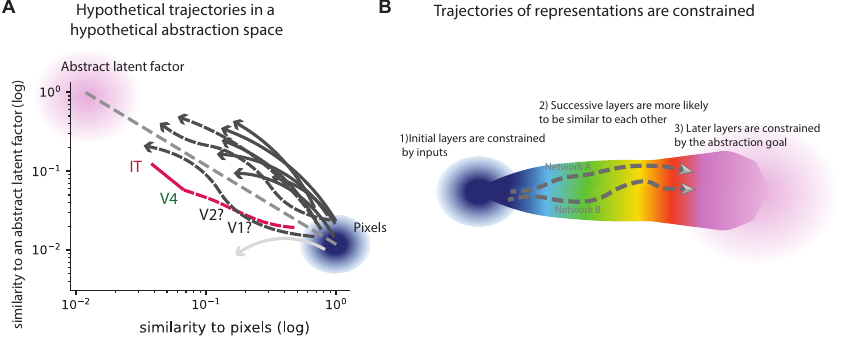}
    \caption{\textbf{A)} Each path illustrates how hypothetical deep neural networks (DNNs) and neural data progress from pixel-level representations (shown as a navy bulb) to more abstract latent factors (shown as a pink bulb).  \textbf{B)} The similarity of intermediate representations between Network A and Network B might be caused by the fact that both networks have to path through similar abstraction stages.}
    \label{fig:fig2}
\end{figure}

% \subsection*{Constrained trajectory of representations in abstraction space mediates representational similarity}
\subsection*{Discussion}
Neural networks can be optimized for different objectives related to their input modality (e.g., object recognition for vision, music or speech recognition for audition, and odor recognition for olfaction). In vision, these objectives can include object classification, scene segmentation, object detection, or self-supervised objectives such as contrastive learning or autoencoders, which aim to create abstract representations of image pixels. In the stimulus set we studied, object classes represented the most abstract latent factor, underlying the generation of all 5760 images. Although the most abstract latent factor may vary depending on the images in a stimulus set, it remains, in any case, distant from the pixel space in the abstraction coordinate system.\\ 

Neural networks operating on pixels of the stimulus set form representations that are initially (early layers in DNNs, or early timestamps in recurrent neural networks) more similar to pixels, thus being located close to the pixel representation in the abstraction space (Figure \ref{fig:fig2}A). The trajectory endpoint, depending on the objective function (e.g. detection, segmentation, classification), is located away from the pixel space and closer to the abstract latent factor, such as classification. Thus, the spectrum of intermediate representation trajectories is bounded at one end by the input (e.g., the stimulus set’s pixels) and at the other by the latent generative factor (e.g., classes). The abstraction coordinates of successive layers are typically close to each other, indicating that the network's trajectory can be viewed as a gradually evolving continuous path. This holds even though each layer is represented as a discrete point in the abstraction coordinates (see Figure \ref{fig:fig2} B). With these limitations, the space that the trajectories of different networks can traverse is limited, leading to trajectories that are close together in the abstraction coordinates. As observed in this study, proximity in the abstraction coordinates reflects representational similarity, irrespective of the network's objective; networks optimized for different objectives exhibit similarities in their representations across the abstraction hierarchy. \\
% Finally, drawing intuition from the intermediate value theorem, one can infer about the intermediate similarity of two trajectories given the relation between the start and end of trajectories (or their projection on a common axis). The Intermediate Value Theorem states that if two functions $f(x)$ and $g(x)$ are continuous on the interval $[a,b]$ and $f(a) > g(a)$ and $g(b) > f(b)$, then there exists at least one c in the interval (a,b) such that $f(c) = g(c)$ (Figure \ref{fig:fig2}C).

% \section*{Conclusions}
In this work, we explored the representational similarity between biological and artificial neural visual systems by studying them in a new abstraction space. This analysis suggests that the observed similarity between these systems is shaped by the constrained trajectories these representations follow, transitioning from pixel-level to more abstract class-level representations. This understanding suggests inherent constraints in the evolution of neural representations and provides evidence for the utility of the abstraction space as a useful framework for analyzing this evolution. The correlation between spatial proximity in abstraction space and representational similarity showcases its potential to further explore the links between biological and artificial visual systems. This study invites further investigation to deepen our understanding of the inherent constraints on neural representations in task-optimized systems, potentially bridging insights between neuroscience and artificial intelligence in understanding visual recognition.

% By drawing parallels with the Intermediate Value Theorem, we have provided a glimpse into the potential mathematical foundations governing the similarities between representational trajectories.

% Before the advancement in deep learning models of object recognition marked by \citep{krizhevsky14} which trailblazing the possibility of measuring the similarity of biological and artificial object recognition systems, neurophysiologists had shown that in the ventral visual cortex of primates, the population of neurons respond to low-level features such as edges in earlier areas (V1) and to complex objects and object categories in later layers such Inferotemporal (IT) cortex and prefrontal cortex.

\clearpage
\section*{Acknowledgment}
I would like to thank the DiCarlo lab at MIT for providing the neural data. T.T. is supported by NIH 1K99EY035357-01. This work was also supported by NIH RF1DA056397, NSF 1707398, and Gatsby Charitable Foundation GAT3708.

\bibliography{ref}
%%%%%%%%%%%%%%%%%%%%%%%%%%%%%%%%%%%%%%%%%%%%%%%%%%%%%%%%%%%%

\end{document}